\documentclass[reprint,amsmath,amssymb,aps]{revtex4-1}
\usepackage{graphicx}
\usepackage{dcolumn}
\usepackage{bm}
\usepackage[normalem]{ulem}
\usepackage{color}
\usepackage{float}

\begin{document}

\title{Fourier Amplitude Distribution and Intermittency in Mechanically Generated Surface Gravity Waves}

    \author{Elmira Fadaeiazar}
    \email{elmira.fadaeiazar@gmail.com}
    \author{Justin Leontini}
    \affiliation{Department of Mechanical and Product Design Engineering, Swinburne University of Technology, 3122, Hawthorn, VIC, Australia}
    \author{Miguel Onorato}
    \affiliation{Dipartimento di Fisica, Universit{\'a} di Torino, 10125, Torino, Italy}
    \affiliation{INFN, 10125 Torino, Italy}
    \author{Takuji Waseda}
    \affiliation{Department of Ocean Technology, Policy and Environment, Graduate School of Frontier Sciences, The University of Tokyo, Tokyo, Japan}
    \author{Alberto Alberello}
    \email{alberto.alberello@outlook.com}
    \affiliation{School of Mathematical Sciences, University of Adelaide, 5005, Adelaide, SA, Australia}
    \author{Alessandro Toffoli}
    \affiliation{Department of Infrastructure Engineering, The University of Melbourne, 3010, Parkville, VIC, Australia}
    
\date{\today}

\begin{abstract}

We examine and discuss the spatial evolution of the statistical properties of mechanically generated surface gravity wave fields, initialised with unidirectional spectral energy distributions, uniformly distributed phases and Rayleigh distributed amplitudes. We demonstrate that nonlinear interactions produce an energy cascade towards high frequency modes with a directional spread and triggers localised intermittent bursts. By analysing the probability density function of Fourier mode amplitudes in the high frequency range of the wave energy spectrum, we show that a heavy-tailed distribution emerges with distance from the wave generator as a result of these intermittent bursts, departing from the originally imposed Rayleigh distribution, even under relatively weak nonlinear conditions.

\end{abstract}

\pacs{Valid PACS appear here}
\keywords{intermittency, wave turbulence, waves, weak wave turbulence}
\maketitle

\section {Introduction}

Ocean waves are a random process.
In the ideal condition of infinite water depth, infinitesimally small wave amplitudes and narrow-banded wave energy spectrum, linear mechanisms dominate wave physics at small time scales. Consequently, the surface elevation can be described as a sum of a large number of harmonics, each with amplitude randomly chosen from a Rayleigh distribution and phases uniformly distributed. Under these conditions, instantaneous displacements of the surface elevation follow a Gaussian distribution \cite{longuet1952statisticaldistribution,ochi2005ocean}.

In nature, however, the small amplitude assumption does not hold.
As a result, mutual  interactions between wave components occur, giving rise to second order nonlinear effects that force the emergence of a weakly non-Gaussian regime \cite{tayfun1980narrow,forristall2000wave,toffoli2007second,mcallister2019jfm}. If waves are sufficiently energetic and their energy spectrum is narrow banded (i.e.~energy is concentrated around a dominant mode), higher order nonlinearity can further develop, triggering the formation of large amplitude waves, also known as rogue waves, and concurrently imposing strong deviations from Gaussian statistics to the surface elevation \cite{onorato01,janssen03,koro2008,onorato2009prl,onorato2009statistical,chabchoub2013prl}.
Although laboratory experiments and numerical simulations have demonstrated that there are sea states particularly prone to extreme events \cite{onorato2009statistical,waseda2009evolution,toffoli2010evolution,mcallister2019jfm}, field observations of rogue waves are serendipitous and a weakly non-Gaussian statistics is normally considered sufficient to describe oceanic wave fields \cite{forristall2000wave,fedele2016real}. 

High order nonlinearity is also responsible for a redistribution of energy and wave action across modes, which modifies the shape of the wave energy spectrum \cite{onorato2009statistical,waseda2009evolution,koro2008}. 
In this respect, a fraction of wave energy moves to low frequency components, producing a downshift of the spectral peak (the dominant wave component becomes longer). Another fraction cascades towards high frequencies, forming an equilibrium tail in the wave spectrum that decays as $\omega^{-4}$ \cite{zakharov2012kolmogorov,onorato2002freely,toffoli2017wind}, with $\omega$ being the angular wave frequency.
Note, however, that wave breaking can force the equilibrium tail to shift toward $\omega^{-5}$ \citep{romero2012jpo,lenain2017jpo}.
In the laboratory, furthermore, mechanically generated waves can exhibit a spectral tail $\omega^{-5}$ or even steeper \citep{deike2014pre,deike2015role,onorato2009statistical,waseda2009interplay,FadaeiazarElmira2018Wtai} due to energy damping, also confirmed by numerical simulations \cite{koro2008}.
This energy cascade, nevertheless, resembles the one described by the Kolmogorov-type velocity spectrum in high Reynolds number flows and it is normally referred to as weak wave turbulence \cite{zakharov2012kolmogorov}. Interestingly, according to the weak wave turbulence theory, a direct cascade characterized by constant flux of energy is not possible in 1D and infinite water depth because the coupling coefficient in the wave kinetic equation is identically zero on the resonant manifold \cite{dyachenko1994free}, i.e.~nonlinear interactions that form the energy cascade at the spectral tail are intrinsically 2D. Therefore, the energy cascade also redistributes on multiple directions, forcing the directional spreading of an initially unidirectional spectral energy distribution or the broadening of an already directional wave spectrum \cite{onorato2009statistical,toffoli10}.

        \begin{figure}
            \centerline{\includegraphics[width=1\linewidth]{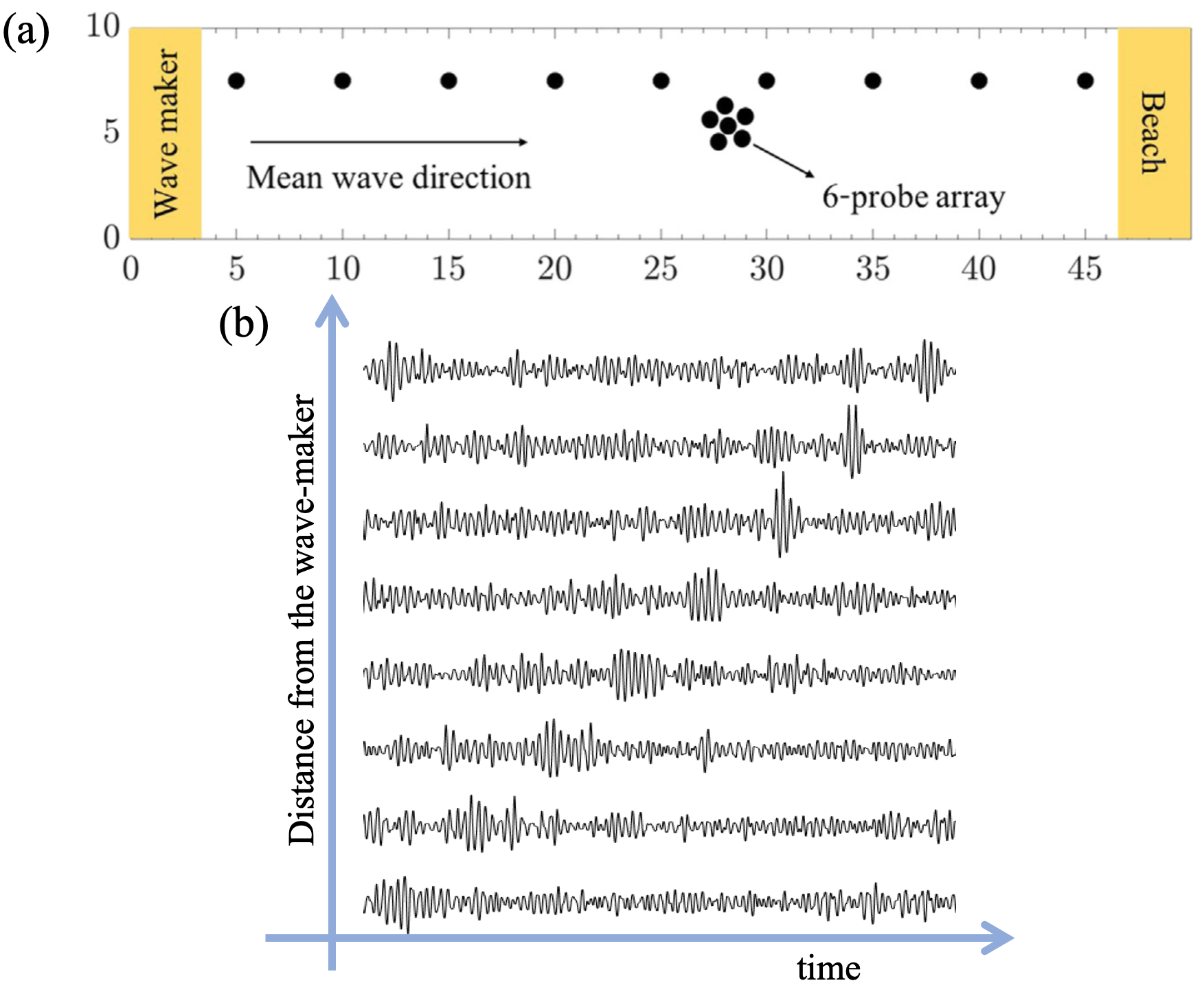}}
            \caption{\label{fig:pos}Schematic (not in scale) of the Ocean Engineering Tank at The University of Tokyo (panel a); sample time series along the tank (panel b).}
         \end{figure}

Similarly to classical hydrodynamic turbulence, weak wave turbulence can exhibit non-Gaussian bursts in the surface elevation---a phenomenon known as intermittency \cite{denissenko2007prl,falcon2007observation,falcon2010origin,nazarenko2010statistics,randoux2014intermittency,FadaeiazarElmira2018Wtai}---that causes a deviation from classical predictions.
These bursts are associated with, but not necessarily limited to, the presence of coherent structures on the water surface and propagating breaking waves \cite{connaughton2003dimensional,yokoyama2004statistics,choi2005anomalous,falcon2010origin} and they are caused by intense nonlinear interactions \cite{deike2015role}. Note that bursts are still conspicuous, yet less pronounced, in directional sea states, where the level nonlinearity is weaker \cite{FadaeiazarElmira2018Wtai}.

As foreshadowed in \cite{denissenko2007prl,nazarenko2010statistics}, intermittent bursts cause mode amplitudes to naturally depart from the hypothesised Rayleigh distribution, especially for the high frequency modes. A systematic analysis of the effect of intermittency on the statistical properties of mode amplitudes, however, has not yet been carried out.  
Understanding wave properties at small scales, nonetheless, is crucial for characterising ocean surface roughness, which is of interest in a wide range of research areas including wind wave generation, air-sea interaction, gas exchange, and ocean remote sensing (e.g. \cite{hwang2013ocean}). 

In the present paper, we examine and discuss past experimental data. With respect to previous work, we track the spatial evolution of the statistical properties of initially unidirectional mechanically generated wave fields, to asses the role of different degrees of nonlinearity on intermittency.
After a brief description of the laboratory tests, we discuss general statistical and spectral properties of the surface elevation, demonstrating that weak wave turbulence is responsible for a directionally spread energy cascade and a transition from Rayleigh to non-Rayleigh statistics in the high frequency modes, even under relatively weak nonlinear conditions. The signature of intermittency in the upper tail of the wave spectrum is discussed. 

\section{The experimental model}

An experimental model to track the nonlinear evolution of irregular wave fields was set up in the Ocean Engineering Tank of the Institute of Industrial Science, The University of Tokyo (description of the experiments and additional details can be found in \citep{waseda2015third,toffoli2015rogue,FadaeiazarElmira2018Wtai}). The facility is 10\,m wide and 50\,m long with a water depth of 5\,m. At one end, the tank is equipped with a wave-maker with 32 digitally controlled triangular plungers, while a sloping beach absorbs the incoming wave energy at the opposite end (see schematic in Fig.~\ref{fig:pos}a).

The water surface elevation was measured using nine capacitance wave gauges deployed along the tank at 5\,m intervals (probes were 2.5\,m from the side wall, Fig. \ref{fig:pos}a). An additional array of six probes was placed 27\,m from the wave-maker. The probes were operated at a sampling frequency of 100\,Hz. 

Water waves were generated by the oscillatory motion of the wave-maker. The plungers were forced by a predefined voltage computed using input Fourier amplitudes with modulus randomly chosen from a Rayleigh distribution around a target spectrum and phases randomly chosen from a uniform distribution in $(0,2\pi]$. The input (target) frequency spectrum at the wave maker was defined using a JONSWAP formulation \citep{komen94}:
        \begin{equation}
            S(\omega)=\frac{\alpha g^2}{\omega^5}\exp\left[-\frac{5}{4}\left(\frac{\omega}{\omega_p}\right)^{-4}\right]\gamma^{\exp[(\omega - \omega _p)^2/2\sigma_j^2\omega_p^2]}
        \end{equation}
where $g$ is the acceleration due to gravity,
$\omega_P$ the angular frequency of the spectral peak, $\sigma$ a constant equal to $0.07$ for $\omega \leqslant \omega_p$ and $0.09$ for $\omega > \omega_p$, $\gamma$ the peak enhancement factor and $\alpha$ the Phillip's constant. No directional distribution was applied to the input wave spectrum, i.e.~the initial wave field produced by the wave-maker is unidirectional.

The input spectrum was discretised into 1024 equal energy bins, with angular frequency in the range 0--16\,rad/s; the upper limit is imposed by the mechanical constraints of the wave-maker. A peak (dominant) wave period $T_p=0.8$\,s ($\omega_p=7.85$\,rad/s) was set as initial condition. This corresponds to a peak wavelength $L_p=1$\,m, which allows a spatial evolution of 50 wavelengths along the tank. The peak enhancement factor $\gamma$ dictates the bandwidth of the spectrum and it was selected to be equal to 3, a typical value for ocean waves \citep{komen94}.
The Phillip's constant $\alpha$ controls the amount of energy in the wave field and it was chosen to define two different values of wave steepness $\varepsilon = k_p H_S /2= 0.06$ and 0.11, where $k_p$ is the wavenumber at the spectral peak (computed via the linear dispersion relation, i.e.~$\omega^2=gk$) and $H_S=4\sqrt{m_0}$ is the significant wave height, with $m_0$ the zero-th order moment of the wave spectrum $S(\omega)$. The ratio of steepness to bandwidth is proportional to the Benjamin-Feir Index ($BFI$, \citep{onorato01,janssen03}), i.e.~a measure of the relative significance of wave nonlinearity to dispersion and an indicator for the appearance of extreme/rogue waves (waves with height larger than $2H_S$) when $BFI = \mathcal{O}(1)$. For sea states discussed herein, $BFI=0.4$ and 0.8 for low ($\varepsilon=0.06$) and high ($\varepsilon=0.11$) steepness, respectively. The former refers to a nearly Gaussian random process, the latter to a strongly non-Gaussian system which is more prone to extreme waves \citep{onorato2009statistical,onorato2009prl,waseda2009evolution}. It is worth mentioning that these initial conditions were selected to maximise the occurrence of extreme waves without reaching the breaking onset \cite{toffoli10}. 

For each initial conditions, four 1-hour realisations with different random amplitudes and phases were carried out to ensure enough data points for stable statistical analysis. An example of recorded time series along the tank is reported in Fig.~\ref{fig:pos}b.

        \begin{figure} 
            \centerline{\includegraphics[width=1.1\linewidth]{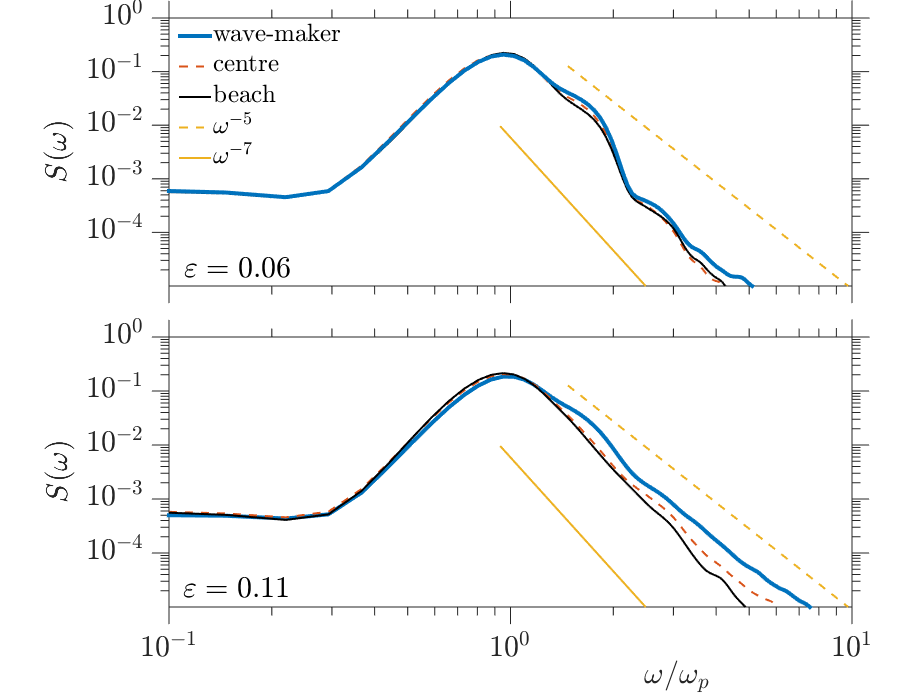}}
            \caption{\label{fig:spec} Wave energy spectra at three positions along the tank (wave-maker, centre and beach) for steepness $\varepsilon=0.06$ ($BFI=0.4$; upper panel) and $\varepsilon=0.11$ ($BFI=0.8$; bottom panel) and reference slopes $\omega^{-7}$ and $\omega^{-5}$.}
        \end{figure}
        
        \begin{figure} 
            \centerline{\includegraphics[width=.95\linewidth]{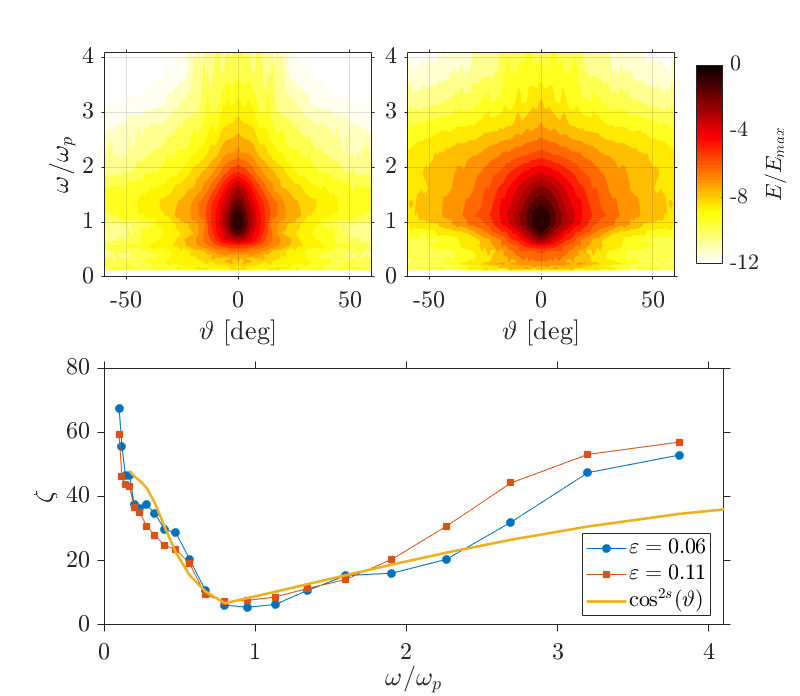}}
            \caption{\label{fig:spec2d} Directional wave energy spectra at the centre of the tank for steepness $\varepsilon=0.06$ ($BFI=0.4$, upper left) and $\varepsilon=0.11$ ($BFI=0.8$, upper right) shown in logarithmic scale and directional spreading as a function of dimensionless wave frequency (bottom). Reference directional distribution $\cos^{2s}(\vartheta)$ is also shown.}
        \end{figure}

\section{Properties of the surface elevation}

\subsection{Energy spectrum}

The input wave spectrum exhibits a drop off for frequencies greater than $2\omega_p$, as these modes are not resolved by the wave-maker. Nevertheless, the spectral tail forms naturally within 5 wavelengths (see details in \cite{FadaeiazarElmira2018Wtai}). As waves propagate farther, the spectrum rapidly evolves due to a nonlinear energy transfer among modes.
The frequency wave spectra recorded along the tank (at 5 wavelengths from the wave-maker, labeled ``wave-maker''; at 25 wavelengths, labeled ``centre''; and at 45 wavelengths, labeled ``beach''), are shown in Fig.~\ref{fig:spec} for both initial conditions; spectral evolution is discussed in more details in \cite{toffoli2015rogue,waseda2015third}.

A fraction of energy is transferred towards low frequencies, resulting in a notable downshift of the spectral peak \cite{onorato2009statistical,waseda2009evolution}. By the centre of the tank, the peak angular frequency $\omega_p$ reduced by approximately 5\%. Note that this apparent spectral downshift is primarily due to detuned resonance as four-wave resonance is effective only when directional spreading of wave energy is broad \cite{waseda2009interplay}. Furthermore, energy cascades towards high-frequency modes soon after the wave generation, forming a complete upper tail within the first five wavelengths of propagation. For the less nonlinear wave field ($\varepsilon=0.06$; $BFI=0.4$), the spectral slope assumes a shape proportional to a power law of $\omega^{-6.1}$, as evaluated with a least square method over the interval $1.5<\omega/\omega_p<8.5$. As waves evolve further, the spectral slope does not change significantly (the tail is proportional to $\omega^{-6.8}$ at the centre and $\omega^{-6.7}$ nearby the beach). 
For the more nonlinear condition ($\varepsilon=0.11$; $BFI=0.8$), the spectral tail shapes as $\omega^{-5.4}$ nearby the wave-maker and further evolve into $\omega^{-6.0}$ at the centre of the tank and $\omega^{-6.6}$ close to the beach. Dependence of spectral slope on the nonlinear properties are consistent with other experimental studies in \citep{onorato2009statistical,waseda2009evolution,waseda2009interplay}.

Although the input wave spectrum imposed at the wave-maker is unidirectional, the width of the tank allows nonlinearity to redistribute energy along two primary directions of $\pm35.5^{\circ}$ with respect to the mean direction of propagation \cite{toffoli2010jgr}, while cascading to high frequencies. This induces the development of a directional distribution (c.f. \cite{onorato2009statistical,waseda2009evolution}), switching the initial frequency-dependent spectral density $S(\omega)$ into a directional spectrum $E(\omega,\vartheta) = S(\omega) \times D(\omega,\vartheta)$, where $D(\omega,\vartheta)$ is a directional spreading function. For the present experiment, $E(\omega,\vartheta)$ was evaluated at the centre of the tank with a wavelet directional method \cite{donelan1996jpo}, using records from the 6-probe array. The reconstructed spectra from both wave fields are presented in the upper panels of Fig. \ref{fig:spec2d} (energy is shown in logarithmic scale to highlight directional spreading at high frequencies). 
Remarkably, the figure confirms that the initial unidirectional wave fields develop into weakly directional ones by the centre of the wave tank. The directional width at each frequency can be summarised as the standard deviation of directional function $D(\omega,\vartheta)$ \citep{holthuijsen2010waves}:
\begin{equation}
    \zeta(\omega) = \sqrt{\int_{-\pi}^{+\pi}\left[2\sin{\frac{\vartheta}{2}}\right]^2 D(\omega,\vartheta)d\vartheta}
\end{equation}
The distribution of $\zeta$ as a function of frequency is reported in the lower panel of Fig. \ref{fig:spec2d}. Directional spreading is narrow at the peak with $\zeta \approx 10^{\circ}$, while it broadens towards lower and higher frequencies ($\zeta \approx 50^{\circ}$). The directional width is marginally more pronounced for the most nonlinear sea state.  

As a reference, the directional width associated to the empirical directional function of the form $cos^{2s}(\vartheta/2)$ proposed by \cite{mitsuyasu1975jpo}---often used to describe directional sea states for engineering applications \cite{goda2010}---is reported in the lower panel of Fig. \ref{fig:spec2d}. 
The frequency-dependent directional coefficient assumes the form $s=(\omega/\omega_p)^{-2.5}s_p$ for $\omega\geq\omega_p$ and $s=(\omega/\omega_p)^{5}s_p$ for $\omega<\omega_p$; the variable $s_p$ is the spreading coefficient at the spectral peak. In Fig. \ref{fig:spec2d}, the spreading of the $cos^{2s}$ function is evaluated with $s_p=25$, which corresponds to a narrow directional swell in the ocean \cite{goda2010}. Interestingly, the directional distribution that develops in the tank is consistent with the empirical counterpart. It is worth noting, however, that the transition from the narrow peak to broad tails is faster in the tank than predicted by \cite{mitsuyasu1975jpo}.

        \begin{figure} 
            \centerline{\includegraphics[width=.9\linewidth]{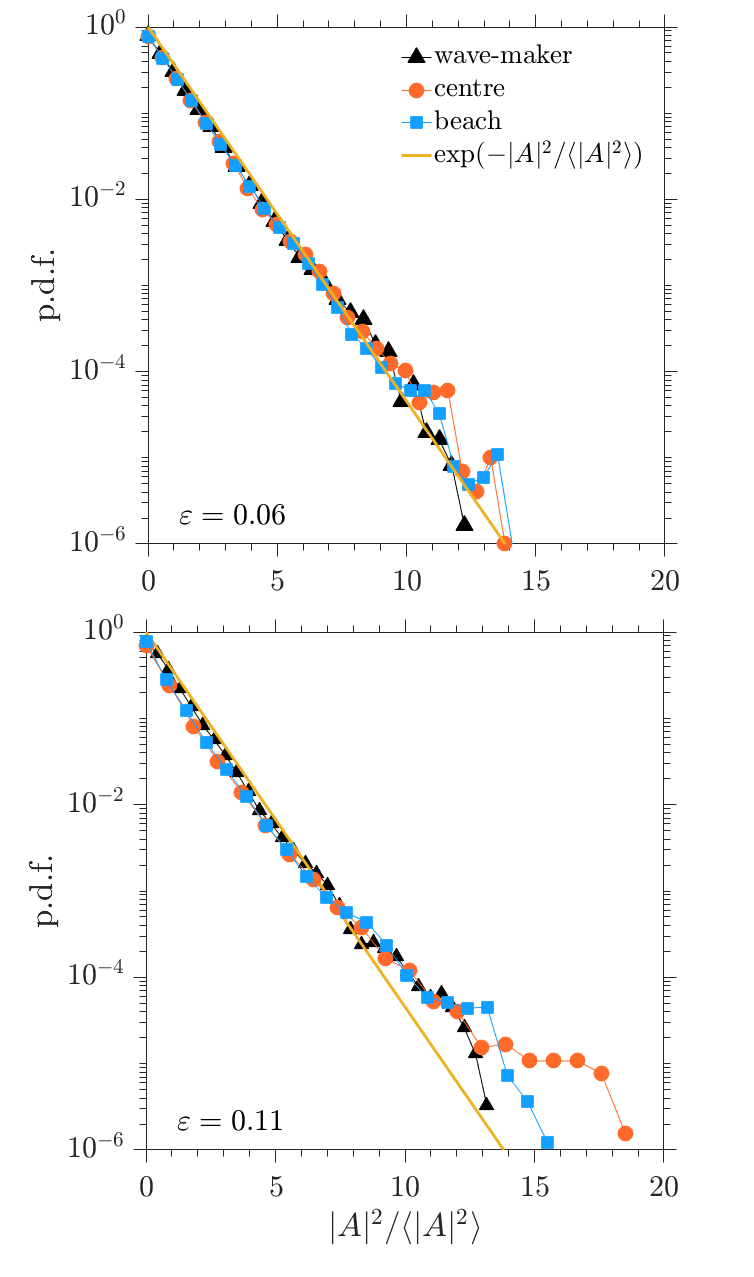}}
            \caption{\label{fig:waveheight}Evolution of the probability density function ($p.d.f.$) of the normalised wave intensity $|A|^2/\langle |A|^2 \rangle$ along the tank: $\varepsilon=0.06$ (upper panel); and $\varepsilon=0.11$ (lower panel). The reference distribution $\exp(-|A|^2/\langle |A|^2 \rangle)$ is also shown.}
        \end{figure} 
        
       \begin{figure*} 
            \centerline{\includegraphics[width=1.1\linewidth]{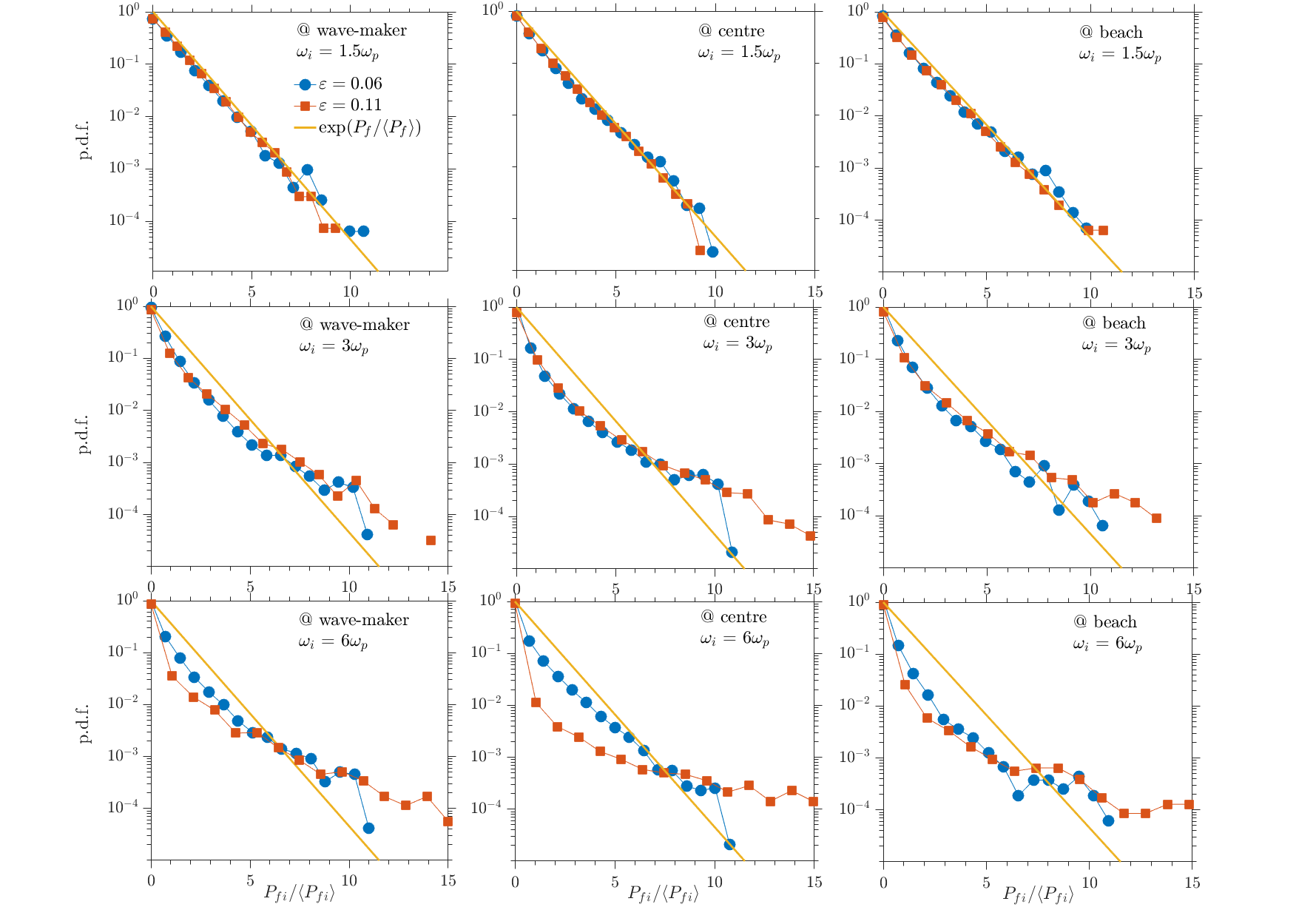}}
            \caption{\label{fig:pdf_amp}Probability density function of the normalised mode intensity $P_f{_i}/\langle P_f{_i} \rangle$ for: $\varepsilon=0.06$ and $BFI=0.4$ (blue circles); and $\varepsilon=0.11$ and $BFI=0.8$ (orange squares). The yellow solid line refers to an exponential distribution.}
        \end{figure*}        
        
\subsection{Wave statistics}

The input wave fields are Gaussian random processes (surface elevation follows a normal distribution, while amplitudes are distributed according to a Rayleigh distribution). Whereas this initial condition remains invariant along the tank if nonlinear properties are weak, non-Gaussian statistics develop for $BFI = \mathcal{O}(1)$ \cite{onorato2009statistical,onorato2009prl,waseda2009evolution}.
For the purpose of the present study, wave statistics of the surface elevation are evaluated by the probability density function ($p.d.f.$) of the wave intensity, defined as the square modulus of the wave envelope $|A^2|$ \cite{janssen2014random}; the average intensity $\langle |A^2| \rangle$ is applied as normalising factor. For a Gaussian random process, the intensity follows an exponential function, which represents a benchmark for the Rayleigh distribution \cite{janssen2014random,toffoli2017wind,el2018spontaneous}. Fig.~\ref{fig:waveheight} shows the $p.d.f.$ of wave intensity at different distances from the wave-maker (nearby the wave-maker, at the centre of the tank, and nearby the beach) and for both initial conditions.

For the less nonlinear sea state ($\varepsilon = 0.06$; $BFI = 0.4$), the intensity is consistent with the exponential distribution throughout the tank (upper panel in Fig.~\ref{fig:waveheight}), indicating that nonlinear interactions, if any, do not affect the statistical properties of the surface elevation, which remains a Gaussian random process. Note that the wave envelope does not capture second-order effects and the concurrent weak deviations from Gaussianity (e.g. \cite{toffoli2007second}). On the contrary, the wave intensity exhibits notable departures from the exponential function for the more nonlinear case ($\varepsilon =0.11$; $BFI = 0.8$). Under these circumstances, nonlinear interactions are responsible for the growth of large individual waves, forcing the emergence of an heavy tail in the $p.d.f.$ Departures from the exponential function enhance with distance from the wave-maker, marking a transition from a weakly to a strongly non-Gaussian process as the wave field evolves in space (lower panel in Fig.~\ref{fig:waveheight}).
A similar behavior has been observed in other experiments tracking wave statistics in water and nonlinear fibers (e.g.~\cite{onorato2009statistical,onorato2009prl,waseda2009evolution,el2018spontaneous}, among others). Although the significant wave height remains unchanged throughout the tank \cite{toffoli2015rogue,waseda2015third} and breaking probability is minimised for the selected sea state conditions \cite{toffoli10}, it cannot be excluded that sporadic wave breaking might have affected the highest waves in the records \cite{toffoli10,alberello2019observation} and hence the $p.d.f.$ at low probability levels.

\section{Single Fourier Mode Analysis}

\subsection{Probability density function of mode intensity}

Time series were subdivided into consecutive blocks of 256 data points with 50\% overlap and a Fourier Transform with Hanning window was applied to convert these blocks into frequency amplitude spectra. A record for statistical analysis was compiled by including Fourier mode amplitudes $A_{fi}$ from all the blocks at frequencies $0.5\omega_p \leqslant \omega_i \leqslant 6\omega_p$.
At each mode, the corresponding intensity ($P_{fi}$) was computed as the modulus of the squared amplitudes (i.e.~$|A_{fi}^2|$); for convenience, intensities were normalised by the concurrent mean $\langle P_{fi} \rangle$. Consistent with the wave intensity $|A^2|$, $P_{fi}$ also distributes according to an exponential function, which represents a benchmark for Rayleigh distributed modes.

\begin{figure} 
            \centerline{\includegraphics[width=1\linewidth]{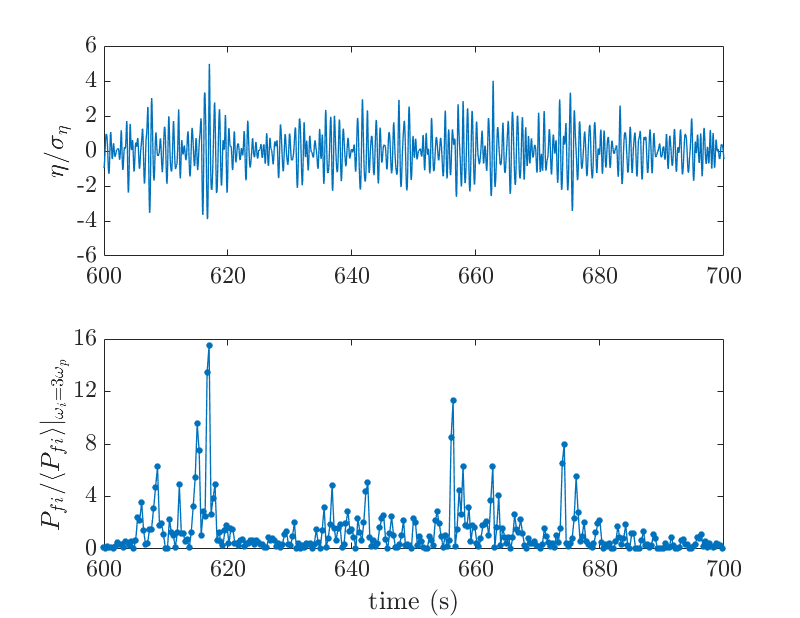}}
            \caption{\label{fig:burst}Example of the surface elevation (top) and intermittent burst at $3\omega_p$ (bottom) for nonlinear wave field ($\varepsilon=0.11$ and $BFI=0.8$).}
        \end{figure}

Fig.~\ref{fig:pdf_amp} shows the $p.d.f.$ of $P_{fi}/\langle P_{fi} \rangle$ for modes at 1.5$\omega_p$, 3$\omega_p$ and 6$\omega_p$ and at different distances from the wave-maker.
Modes close to the spectral peak ($1.5\omega_p$, upper panels in Fig.~\ref{fig:pdf_amp}), distribute according to the exponential function $\exp(-P_{fi}/\langle P_{fi} \rangle)$ at any distances from the wave-maker and regardless of the level of nonlinearity of the wave field, i.e.~Fourier mode amplitudes in the proximity of the spectral peak maintain the initial Rayleigh distribution.
Mode intensities at higher frequencies ($3\omega_p$ and $6\omega_p$) exhibit conspicuous deviations from the exponential function, with departures enhancing with distance from the wave-maker (middle and bottom panels in Fig.~\ref{fig:pdf_amp}).
Specifically, the exponential function over-estimates the probability density for $P_{fi}/\langle P_{fi} \rangle < 7$ and under-estimates for $P_{fi}/\langle P_{fi} \rangle > 7$, i.e.~large intensities happen far more frequently than an exponential distribution would predict.
Whereas there is little difference among sea states at low probability levels (middle panels in Fig~\ref{fig:pdf_amp}), a heavier tail emerges at high probability levels for more nonlinear wave fields ($\varepsilon =0.11$; $BFI = 0.8$). This heavy tail is associated with the presence of intermittent bursts
\cite{nazarenko2010statistics} that are generated as a consequence of the anomalous scaling in the energy cascade \cite{falcon2007observation,denissenko2007prl,nazarenko2010statistics}, symptomatic of deviations from the classical weak wave turbulence regime.
An example of the surface elevation $\eta$ (normalised by its standard deviation $\sigma_{\eta}$) and the concurrent non-dimensional intensities $P_{fi}/\langle P_{fi} \rangle$ at $\omega=3 \omega_p$ for the nonlinear case $\varepsilon=0.11$ and $BFI=0.8$ are presented in Fig.~\ref{fig:burst}. Bursts of high intensity equal to or greater than 10 times the standard deviation of the modal surface elevation are evident and often associated to large wave groups. 
        
Although divergence from the exponential function is qualitatively similar at 3$\omega_p$ and 6$\omega_p$,
departures are far more pronounced at $6\omega_p$, indicating the development of strongly non-Rayleigh distributed properties at modes in the upper range of the spectral tail (bottom panels in Fig~\ref{fig:pdf_amp}).
It is worth noting that the $p.d.f.$ of high frequency modes shows a notable dependence on distance from the wave-maker, especially for the more nonlinear sea state ($\varepsilon=0.11$; $BFI=0.8$). There is an evident evolution strengthening the non-Rayleigh properties as waves propagate along the tank and energy cascades to high frequencies. For the less nonlinear case ($\varepsilon=0.06$; $BFI=0.4$) this evolution is still notable but slower, consistent with the weaker nonlinearity of the wave field (cf. \cite{onorato2009statistical}).
The observed dependence of the $p.d.f.$ on distance from the wave-maker is an example of how nonlinear effects take time to develop, and how this timescale depends on wave steepness.

\subsection{Kurtosis of Fourier mode amplitudes}

\begin{figure*} 
            \centerline{\includegraphics[width=\linewidth]{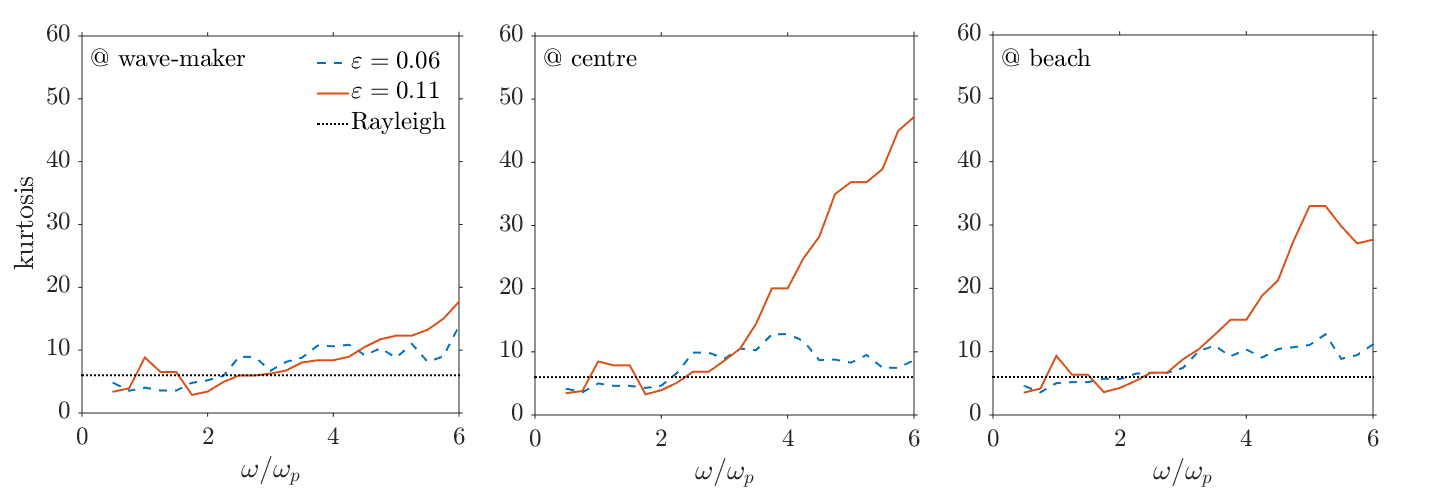}}
            \caption{\label{fig:ku}Kurtosis of Fourier mode amplitudes $A_{fi}$ for the entire frequency range: weakly nonlinear sea state $\varepsilon=0.06$ (blue dashed line); nonlinear sea state $\varepsilon=0.11$ (red solid line); benchmark value of kurtosis for Rayleigh distributed Fourier modes (black dotted line).}
        \end{figure*}

Deviations from the underlying Rayleigh distribution can be summarised over the entire frequency band by the fourth-order statistical moment (i.e.~kurtosis) of the Fourier mode amplitudes $A_{fi}$. The kurtosis describes the tailedness of the distribution and it is equal to 6 for a Rayleigh distributed population. Fig.~\ref{fig:ku} shows the kurtosis as a function of frequency at different distances from the wave-maker and for both sea states.
        
For modes around the spectral peak, approximately $0.5\omega_p\le\omega_i\leq2.5\omega_p$, amplitudes follow the Rayleigh distribution. Observed values of kurtosis are comparable with the reference value of 6, regardless of the degree of nonlinearity of the sea state and the distance from the wave-maker. This result is consistent with the exponential distribution of intensity $P_{fi}$ observed for modes at 1.5$\omega_p$ (upper panels in Fig.~\ref{fig:pdf_amp}).

Modes at high frequencies ($\omega_i > 2.5\omega_p$) exhibit kurtosis greater than 6 for both sea states, indicating the emergence of intermittency in the spectral tail and thus deviations from the classical weak wave turbulence regime.
Near the wave-maker, deviations from the Rayleigh distribution are weak as nonlinearity requires some space to fully develop.
Nevertheless, deviations do not enhance notably for the weak sea state ($\varepsilon=0.06$; $BFI=0.4$) as the wave field moves towards the beach. Although intermittent bursts appear in this wave condition too, their impact on the statistical properties of high frequency modes remains small (this is also consistent with the weak deviation at the tail of the $p.d.f.$ of $P_{fi}$). On the contrary, energy cascading at high frequency gives rise to a substantial divergence from the Rayleigh distribution for more nonlinear sea states ($\varepsilon=0.11$; $BFI=0.8$).
The maximum departure is at the centre of the tank (approximately 25 wavelengths from the wave-maker) where kurtosis reaches values well above 20 for $\omega_i > 4 \omega_p$, in what appears a monotonic increase of kurtosis with frequency. Towards the end of the tank, sporadic wave breaking partially affects high frequency modes, resulting in a drop of the kurtosis. Exceptionally large values of kurtosis reflects the substantial deviation of $P_{fi}$ at the tail of the $p.d.f.$ for modes at $3\omega_p$ and $6\omega_p$ (cf. Fig.~\ref{fig:pdf_amp}).   

\section {Conclusions}

An experimental model to track the spatial evolution of random wave fields has been investigated in the framework of weak wave turbulence. Tests were carried out at the Ocean Engineering Tank of the University of Tokyo. Waves were mechanically generated at one end of the tank to replicate sea states with two target spectral configurations: a weakly nonlinear ($\varepsilon=0.06$; $BFI=0.4$) and a nonlinear sea state ($\varepsilon=0.11$; $BFI=0.8$). 
Conditions at the wave-maker were initialised using Rayleigh distributed amplitudes around a unidirectional target spectrum and uniformly distributed phases, i.e.~the initial conditions were a Gaussian random processes.

As the wave fields propagated along the tank, nonlinear interactions freely developed, producing an energy transfer across modes. The macroscopic effect was the growth of large amplitude waves, which modified the initial statistical properties of the wave fields. For the weakly nonlinear sea state, statistics of the intensity of the surface elevation were Gaussian. For the more nonlinear condition, on the other hand, large amplitude waves were observed far more often than a Gaussian distribution would have predicted, producing heavy tails in the probability density function of wave intensity.
 
Nonlinear energy transfer, moreover, produced changes to the spectral space. A notable fraction of energy moved towards low frequency modes and the spectral peak downshifted. Concomitantly, nonlinear wave interaction transferred energy to high frequencies and redistributed it on multiple directions, causing the initial unidirectional wave field to evolved into a weakly directional one. Note that emergence of directionality was more pronounced for the nonlinear case ($\varepsilon=0.11$; $BFI=0.8$).

Energy cascading towards high frequency---as in a classical fluid turbulent regime---was associated with the appearance of intermittent bursts localised at high frequency modes. This resulted in the emergence of a heavy-tailed distribution for Fourier mode amplitudes, departing from the originally imposed Rayleigh distribution. Deviations were evident in both sea states, albeit they were less pronounced in the weakly nonlinear field ($\varepsilon=0.06$; $BFI=0.4$) and more substantial for nonlinearity sea state ($\varepsilon=0.11$; $BFI=0.8$).
These results provide evidence that the general hypothesis that Fourier mode amplitudes are Rayleigh distributed does not extend to the high frequency range (approximately $\omega_i > 2.5 \omega_p$), even for weakly nonlinear (almost linear) sea state conditions.   

{\bf Acknowledgments}
This research was supported by the JSPS Fellowship for Research in Japan Program, Grants-in-Aid for Scientific Research of the Japanese Society for for the Promotion of Science and the International Science Linkages (ISL) Program of the Australian Academy of Science. M.O. received support from  Simons Collaboration on Wave Turbulence, Award 617006 and from ``Departments of Excellence 2018--2022'' Grant awarded by the Italian Ministry of Education, University and Research (MIUR, L.232/2016). A.A. and A.T. acknowledge support from the Air-Sea-Ice Lab Project. M.O. is grateful to Dr. B. GiuliNico for discussions.

    \bibliography {References}

\begin{thebibliography}{46}%
\makeatletter
\providecommand \@ifxundefined [1]{%
 \@ifx{#1\undefined}
}%
\providecommand \@ifnum [1]{%
 \ifnum #1\expandafter \@firstoftwo
 \else \expandafter \@secondoftwo
 \fi
}%
\providecommand \@ifx [1]{%
 \ifx #1\expandafter \@firstoftwo
 \else \expandafter \@secondoftwo
 \fi
}%
\providecommand \natexlab [1]{#1}%
\providecommand \enquote  [1]{``#1''}%
\providecommand \bibnamefont  [1]{#1}%
\providecommand \bibfnamefont [1]{#1}%
\providecommand \citenamefont [1]{#1}%
\providecommand \href@noop [0]{\@secondoftwo}%
\providecommand \href [0]{\begingroup \@sanitize@url \@href}%
\providecommand \@href[1]{\@@startlink{#1}\@@href}%
\providecommand \@@href[1]{\endgroup#1\@@endlink}%
\providecommand \@sanitize@url [0]{\catcode `\\12\catcode `\$12\catcode
  `\&12\catcode `\#12\catcode `\^12\catcode `\_12\catcode `\%12\relax}%
\providecommand \@@startlink[1]{}%
\providecommand \@@endlink[0]{}%
\providecommand \url  [0]{\begingroup\@sanitize@url \@url }%
\providecommand \@url [1]{\endgroup\@href {#1}{\urlprefix }}%
\providecommand \urlprefix  [0]{URL }%
\providecommand \Eprint [0]{\href }%
\providecommand \doibase [0]{http://dx.doi.org/}%
\providecommand \selectlanguage [0]{\@gobble}%
\providecommand \bibinfo  [0]{\@secondoftwo}%
\providecommand \bibfield  [0]{\@secondoftwo}%
\providecommand \translation [1]{[#1]}%
\providecommand \BibitemOpen [0]{}%
\providecommand \bibitemStop [0]{}%
\providecommand \bibitemNoStop [0]{.\EOS\space}%
\providecommand \EOS [0]{\spacefactor3000\relax}%
\providecommand \BibitemShut  [1]{\csname bibitem#1\endcsname}%
\let\auto@bib@innerbib\@empty
\bibitem [{\citenamefont
  {Longuet-Higgins}(1952)}]{longuet1952statisticaldistribution}%
  \BibitemOpen
  \bibfield  {author} {\bibinfo {author} {\bibfnamefont {M.~S.}\ \bibnamefont
  {Longuet-Higgins}},\ }\href@noop {} {\bibfield  {journal} {\bibinfo
  {journal} {J. Mar. Res.}\ }\textbf {\bibinfo {volume} {11}},\ \bibinfo
  {pages} {245} (\bibinfo {year} {1952})}\BibitemShut {NoStop}%
\bibitem [{\citenamefont {Ochi}(2005)}]{ochi2005ocean}%
  \BibitemOpen
  \bibfield  {author} {\bibinfo {author} {\bibfnamefont {M.~K.}\ \bibnamefont
  {Ochi}},\ }\href@noop {} {\emph {\bibinfo {title} {Ocean waves: the
  stochastic approach}}},\ Vol.~\bibinfo {volume} {6}\ (\bibinfo  {publisher}
  {Cambridge University Press},\ \bibinfo {year} {2005})\BibitemShut {NoStop}%
\bibitem [{\citenamefont {Tayfun}(1980)}]{tayfun1980narrow}%
  \BibitemOpen
  \bibfield  {author} {\bibinfo {author} {\bibfnamefont {M.~A.}\ \bibnamefont
  {Tayfun}},\ }\href@noop {} {\bibfield  {journal} {\bibinfo  {journal} {J.
  Geophys. Res.}\ }\textbf {\bibinfo {volume} {85}},\ \bibinfo {pages} {1548}
  (\bibinfo {year} {1980})}\BibitemShut {NoStop}%
\bibitem [{\citenamefont {Forristall}(2000)}]{forristall2000wave}%
  \BibitemOpen
  \bibfield  {author} {\bibinfo {author} {\bibfnamefont {G.~Z.}\ \bibnamefont
  {Forristall}},\ }\href@noop {} {\bibfield  {journal} {\bibinfo  {journal} {J.
  Phys. Oceanogr.}\ }\textbf {\bibinfo {volume} {30}},\ \bibinfo {pages} {1931}
  (\bibinfo {year} {2000})}\BibitemShut {NoStop}%
\bibitem [{\citenamefont {Toffoli}\ \emph {et~al.}(2007)\citenamefont
  {Toffoli}, \citenamefont {Monbaliu}, \citenamefont {Onorato}, \citenamefont
  {Osborne}, \citenamefont {Babanin},\ and\ \citenamefont
  {Bitner-Gregersen}}]{toffoli2007second}%
  \BibitemOpen
  \bibfield  {author} {\bibinfo {author} {\bibfnamefont {A.}~\bibnamefont
  {Toffoli}}, \bibinfo {author} {\bibfnamefont {J.}~\bibnamefont {Monbaliu}},
  \bibinfo {author} {\bibfnamefont {M.}~\bibnamefont {Onorato}}, \bibinfo
  {author} {\bibfnamefont {A.}~\bibnamefont {Osborne}}, \bibinfo {author}
  {\bibfnamefont {A.}~\bibnamefont {Babanin}}, \ and\ \bibinfo {author}
  {\bibfnamefont {E.}~\bibnamefont {Bitner-Gregersen}},\ }\href@noop {}
  {\bibfield  {journal} {\bibinfo  {journal} {J. Phys. Oceanogr.}\ }\textbf
  {\bibinfo {volume} {37}},\ \bibinfo {pages} {2726} (\bibinfo {year}
  {2007})}\BibitemShut {NoStop}%
\bibitem [{\citenamefont {McAllister}\ \emph {et~al.}(2019)\citenamefont
  {McAllister}, \citenamefont {Draycott}, \citenamefont {Adcock}, \citenamefont
  {Taylor},\ and\ \citenamefont {van~den Bremer}}]{mcallister2019jfm}%
  \BibitemOpen
  \bibfield  {author} {\bibinfo {author} {\bibfnamefont {M.~L.}\ \bibnamefont
  {McAllister}}, \bibinfo {author} {\bibfnamefont {S.}~\bibnamefont
  {Draycott}}, \bibinfo {author} {\bibfnamefont {T.~A.~A.}\ \bibnamefont
  {Adcock}}, \bibinfo {author} {\bibfnamefont {P.~H.}\ \bibnamefont {Taylor}},
  \ and\ \bibinfo {author} {\bibfnamefont {T.~S.}\ \bibnamefont {van~den
  Bremer}},\ }\href {\doibase 10.1017/jfm.2018.886} {\bibfield  {journal}
  {\bibinfo  {journal} {Journal of Fluid Mechanics}\ }\textbf {\bibinfo
  {volume} {860}},\ \bibinfo {pages} {767–786} (\bibinfo {year}
  {2019})}\BibitemShut {NoStop}%
\bibitem [{\citenamefont {Onorato}\ \emph {et~al.}(2001)\citenamefont
  {Onorato}, \citenamefont {Osborne}, \citenamefont {Serio},\ and\
  \citenamefont {Bertone}}]{onorato01}%
  \BibitemOpen
  \bibfield  {author} {\bibinfo {author} {\bibfnamefont {M.}~\bibnamefont
  {Onorato}}, \bibinfo {author} {\bibfnamefont {A.}~\bibnamefont {Osborne}},
  \bibinfo {author} {\bibfnamefont {M.}~\bibnamefont {Serio}}, \ and\ \bibinfo
  {author} {\bibfnamefont {S.}~\bibnamefont {Bertone}},\ }\href@noop {}
  {\bibfield  {journal} {\bibinfo  {journal} {Phys. Rev. Lett.}\ }\textbf
  {\bibinfo {volume} {86}},\ \bibinfo {pages} {5831} (\bibinfo {year}
  {2001})}\BibitemShut {NoStop}%
\bibitem [{\citenamefont {Janssen}(2003)}]{janssen03}%
  \BibitemOpen
  \bibfield  {author} {\bibinfo {author} {\bibfnamefont {P.~A. E.~M.}\
  \bibnamefont {Janssen}},\ }\href@noop {} {\bibfield  {journal} {\bibinfo
  {journal} {J. Phys. Ocean.}\ }\textbf {\bibinfo {volume} {33}},\ \bibinfo
  {pages} {863} (\bibinfo {year} {2003})}\BibitemShut {NoStop}%
\bibitem [{\citenamefont {Korotkevich}\ \emph {et~al.}(2008)\citenamefont
  {Korotkevich}, \citenamefont {Pushkarev}, \citenamefont {Resio},\ and\
  \citenamefont {Zakharov}}]{koro2008}%
  \BibitemOpen
  \bibfield  {author} {\bibinfo {author} {\bibfnamefont {A.}~\bibnamefont
  {Korotkevich}}, \bibinfo {author} {\bibfnamefont {A.}~\bibnamefont
  {Pushkarev}}, \bibinfo {author} {\bibfnamefont {D.}~\bibnamefont {Resio}}, \
  and\ \bibinfo {author} {\bibfnamefont {V.}~\bibnamefont {Zakharov}},\ }\href
  {\doibase https://doi.org/10.1016/j.euromechflu.2007.08.004} {\bibfield
  {journal} {\bibinfo  {journal} {European Journal of Mechanics - B/Fluids}\
  }\textbf {\bibinfo {volume} {27}},\ \bibinfo {pages} {361 } (\bibinfo {year}
  {2008})}\BibitemShut {NoStop}%
\bibitem [{\citenamefont {Onorato}\ \emph
  {et~al.}(2009{\natexlab{a}})\citenamefont {Onorato}, \citenamefont {Waseda},
  \citenamefont {Toffoli}, \citenamefont {Cavaleri}, \citenamefont {Gramstad},
  \citenamefont {Janssen}, \citenamefont {Kinoshita}, \citenamefont {Monbaliu},
  \citenamefont {Mori}, \citenamefont {Osborne}, \citenamefont {Serio},
  \citenamefont {Stansberg}, \citenamefont {Tamura},\ and\ \citenamefont
  {Trulsen}}]{onorato2009prl}%
  \BibitemOpen
  \bibfield  {author} {\bibinfo {author} {\bibfnamefont {M.}~\bibnamefont
  {Onorato}}, \bibinfo {author} {\bibfnamefont {T.}~\bibnamefont {Waseda}},
  \bibinfo {author} {\bibfnamefont {A.}~\bibnamefont {Toffoli}}, \bibinfo
  {author} {\bibfnamefont {L.}~\bibnamefont {Cavaleri}}, \bibinfo {author}
  {\bibfnamefont {O.}~\bibnamefont {Gramstad}}, \bibinfo {author}
  {\bibfnamefont {P.}~\bibnamefont {Janssen}}, \bibinfo {author} {\bibfnamefont
  {T.}~\bibnamefont {Kinoshita}}, \bibinfo {author} {\bibfnamefont
  {J.}~\bibnamefont {Monbaliu}}, \bibinfo {author} {\bibfnamefont
  {N.}~\bibnamefont {Mori}}, \bibinfo {author} {\bibfnamefont {A.~R.}\
  \bibnamefont {Osborne}}, \bibinfo {author} {\bibfnamefont {M.}~\bibnamefont
  {Serio}}, \bibinfo {author} {\bibfnamefont {C.~T.}\ \bibnamefont
  {Stansberg}}, \bibinfo {author} {\bibfnamefont {H.}~\bibnamefont {Tamura}}, \
  and\ \bibinfo {author} {\bibfnamefont {K.}~\bibnamefont {Trulsen}},\
  }\href@noop {} {\bibfield  {journal} {\bibinfo  {journal} {Phys. Rev. Lett.}\
  }\textbf {\bibinfo {volume} {102}},\ \bibinfo {pages} {114502} (\bibinfo
  {year} {2009}{\natexlab{a}})}\BibitemShut {NoStop}%
\bibitem [{\citenamefont {Onorato}\ \emph
  {et~al.}(2009{\natexlab{b}})\citenamefont {Onorato}, \citenamefont
  {Cavaleri}, \citenamefont {Fouques}, \citenamefont {Gramstad}, \citenamefont
  {Janssen}, \citenamefont {Monbaliu}, \citenamefont {Osborne}, \citenamefont
  {Pakozdi}, \citenamefont {Serio}, \citenamefont {Stansberg}, \citenamefont
  {Toffoli},\ and\ \citenamefont {Trulsen}}]{onorato2009statistical}%
  \BibitemOpen
  \bibfield  {author} {\bibinfo {author} {\bibfnamefont {M.}~\bibnamefont
  {Onorato}}, \bibinfo {author} {\bibfnamefont {L.}~\bibnamefont {Cavaleri}},
  \bibinfo {author} {\bibfnamefont {S.}~\bibnamefont {Fouques}}, \bibinfo
  {author} {\bibfnamefont {O.}~\bibnamefont {Gramstad}}, \bibinfo {author}
  {\bibfnamefont {P.~A. E.~M.}\ \bibnamefont {Janssen}}, \bibinfo {author}
  {\bibfnamefont {J.}~\bibnamefont {Monbaliu}}, \bibinfo {author}
  {\bibfnamefont {A.~R.}\ \bibnamefont {Osborne}}, \bibinfo {author}
  {\bibfnamefont {C.}~\bibnamefont {Pakozdi}}, \bibinfo {author} {\bibfnamefont
  {M.}~\bibnamefont {Serio}}, \bibinfo {author} {\bibfnamefont {C.~T.}\
  \bibnamefont {Stansberg}}, \bibinfo {author} {\bibfnamefont {A.}~\bibnamefont
  {Toffoli}}, \ and\ \bibinfo {author} {\bibfnamefont {K.}~\bibnamefont
  {Trulsen}},\ }\href@noop {} {\bibfield  {journal} {\bibinfo  {journal} {J.
  Fluid Mech.}\ }\textbf {\bibinfo {volume} {627}},\ \bibinfo {pages} {235}
  (\bibinfo {year} {2009}{\natexlab{b}})}\BibitemShut {NoStop}%
\bibitem [{\citenamefont {Chabchoub}\ \emph {et~al.}(2013)\citenamefont
  {Chabchoub}, \citenamefont {Hoffmann}, \citenamefont {Onorato}, \citenamefont
  {Genty}, \citenamefont {Dudley},\ and\ \citenamefont
  {Akhmediev}}]{chabchoub2013prl}%
  \BibitemOpen
  \bibfield  {author} {\bibinfo {author} {\bibfnamefont {A.}~\bibnamefont
  {Chabchoub}}, \bibinfo {author} {\bibfnamefont {N.}~\bibnamefont {Hoffmann}},
  \bibinfo {author} {\bibfnamefont {M.}~\bibnamefont {Onorato}}, \bibinfo
  {author} {\bibfnamefont {G.}~\bibnamefont {Genty}}, \bibinfo {author}
  {\bibfnamefont {J.~M.}\ \bibnamefont {Dudley}}, \ and\ \bibinfo {author}
  {\bibfnamefont {N.}~\bibnamefont {Akhmediev}},\ }\href {\doibase
  10.1103/PhysRevLett.111.054104} {\bibfield  {journal} {\bibinfo  {journal}
  {Phys. Rev. Lett.}\ }\textbf {\bibinfo {volume} {111}},\ \bibinfo {pages}
  {054104} (\bibinfo {year} {2013})}\BibitemShut {NoStop}%
\bibitem [{\citenamefont {Waseda}\ \emph
  {et~al.}(2009{\natexlab{a}})\citenamefont {Waseda}, \citenamefont
  {Kinoshita},\ and\ \citenamefont {Tamura}}]{waseda2009evolution}%
  \BibitemOpen
  \bibfield  {author} {\bibinfo {author} {\bibfnamefont {T.}~\bibnamefont
  {Waseda}}, \bibinfo {author} {\bibfnamefont {T.}~\bibnamefont {Kinoshita}}, \
  and\ \bibinfo {author} {\bibfnamefont {H.}~\bibnamefont {Tamura}},\
  }\href@noop {} {\bibfield  {journal} {\bibinfo  {journal} {Journal of
  Physical Oceanography}\ }\textbf {\bibinfo {volume} {39}},\ \bibinfo {pages}
  {621} (\bibinfo {year} {2009}{\natexlab{a}})}\BibitemShut {NoStop}%
\bibitem [{\citenamefont {Toffoli}\ \emph
  {et~al.}(2010{\natexlab{a}})\citenamefont {Toffoli}, \citenamefont
  {Gramstad}, \citenamefont {Trulsen}, \citenamefont {Monbaliu}, \citenamefont
  {Bitner-Gregersen},\ and\ \citenamefont {Onorato}}]{toffoli2010evolution}%
  \BibitemOpen
  \bibfield  {author} {\bibinfo {author} {\bibfnamefont {A.}~\bibnamefont
  {Toffoli}}, \bibinfo {author} {\bibfnamefont {O.}~\bibnamefont {Gramstad}},
  \bibinfo {author} {\bibfnamefont {K.}~\bibnamefont {Trulsen}}, \bibinfo
  {author} {\bibfnamefont {J.}~\bibnamefont {Monbaliu}}, \bibinfo {author}
  {\bibfnamefont {E.}~\bibnamefont {Bitner-Gregersen}}, \ and\ \bibinfo
  {author} {\bibfnamefont {M.}~\bibnamefont {Onorato}},\ }\href@noop {}
  {\bibfield  {journal} {\bibinfo  {journal} {J. Fluid Mech.}\ }\textbf
  {\bibinfo {volume} {664}},\ \bibinfo {pages} {313} (\bibinfo {year}
  {2010}{\natexlab{a}})}\BibitemShut {NoStop}%
\bibitem [{\citenamefont {Fedele}\ \emph {et~al.}(2016)\citenamefont {Fedele},
  \citenamefont {Brennan}, \citenamefont {De~Le{\'o}n}, \citenamefont
  {Dudley},\ and\ \citenamefont {Dias}}]{fedele2016real}%
  \BibitemOpen
  \bibfield  {author} {\bibinfo {author} {\bibfnamefont {F.}~\bibnamefont
  {Fedele}}, \bibinfo {author} {\bibfnamefont {J.}~\bibnamefont {Brennan}},
  \bibinfo {author} {\bibfnamefont {S.~P.}\ \bibnamefont {De~Le{\'o}n}},
  \bibinfo {author} {\bibfnamefont {J.}~\bibnamefont {Dudley}}, \ and\ \bibinfo
  {author} {\bibfnamefont {F.}~\bibnamefont {Dias}},\ }\href@noop {} {\bibfield
   {journal} {\bibinfo  {journal} {Sci. Rep.}\ }\textbf {\bibinfo {volume}
  {6}},\ \bibinfo {pages} {27715} (\bibinfo {year} {2016})}\BibitemShut
  {NoStop}%
\bibitem [{\citenamefont {Zakharov}\ \emph {et~al.}(2012)\citenamefont
  {Zakharov}, \citenamefont {L'vov},\ and\ \citenamefont
  {Falkovich}}]{zakharov2012kolmogorov}%
  \BibitemOpen
  \bibfield  {author} {\bibinfo {author} {\bibfnamefont {V.}~\bibnamefont
  {Zakharov}}, \bibinfo {author} {\bibfnamefont {V.~S.}\ \bibnamefont {L'vov}},
  \ and\ \bibinfo {author} {\bibfnamefont {G.}~\bibnamefont {Falkovich}},\
  }\href@noop {} {\emph {\bibinfo {title} {Kolmogorov spectra of turbulence I:
  Wave turbulence}}}\ (\bibinfo  {publisher} {Springer Science \& Business
  Media},\ \bibinfo {year} {2012})\BibitemShut {NoStop}%
\bibitem [{\citenamefont {Onorato}\ \emph {et~al.}(2002)\citenamefont
  {Onorato}, \citenamefont {Osborne}, \citenamefont {Serio}, \citenamefont
  {Resio}, \citenamefont {Pushkarev}, \citenamefont {Zakharov},\ and\
  \citenamefont {Brandini}}]{onorato2002freely}%
  \BibitemOpen
  \bibfield  {author} {\bibinfo {author} {\bibfnamefont {M.}~\bibnamefont
  {Onorato}}, \bibinfo {author} {\bibfnamefont {A.~R.}\ \bibnamefont
  {Osborne}}, \bibinfo {author} {\bibfnamefont {M.}~\bibnamefont {Serio}},
  \bibinfo {author} {\bibfnamefont {D.}~\bibnamefont {Resio}}, \bibinfo
  {author} {\bibfnamefont {A.}~\bibnamefont {Pushkarev}}, \bibinfo {author}
  {\bibfnamefont {V.~E.}\ \bibnamefont {Zakharov}}, \ and\ \bibinfo {author}
  {\bibfnamefont {C.}~\bibnamefont {Brandini}},\ }\href@noop {} {\bibfield
  {journal} {\bibinfo  {journal} {Phys. Rev. Lett.}\ }\textbf {\bibinfo
  {volume} {89}},\ \bibinfo {pages} {144501} (\bibinfo {year}
  {2002})}\BibitemShut {NoStop}%
\bibitem [{\citenamefont {Toffoli}\ \emph {et~al.}(2017)\citenamefont
  {Toffoli}, \citenamefont {Proment}, \citenamefont {Salman}, \citenamefont
  {Monbaliu}, \citenamefont {Frascoli}, \citenamefont {Dafilis}, \citenamefont
  {Stramignoni}, \citenamefont {Forza}, \citenamefont {Manfrin},\ and\
  \citenamefont {Onorato}}]{toffoli2017wind}%
  \BibitemOpen
  \bibfield  {author} {\bibinfo {author} {\bibfnamefont {A.}~\bibnamefont
  {Toffoli}}, \bibinfo {author} {\bibfnamefont {D.}~\bibnamefont {Proment}},
  \bibinfo {author} {\bibfnamefont {H.}~\bibnamefont {Salman}}, \bibinfo
  {author} {\bibfnamefont {J.}~\bibnamefont {Monbaliu}}, \bibinfo {author}
  {\bibfnamefont {F.}~\bibnamefont {Frascoli}}, \bibinfo {author}
  {\bibfnamefont {M.}~\bibnamefont {Dafilis}}, \bibinfo {author} {\bibfnamefont
  {E.}~\bibnamefont {Stramignoni}}, \bibinfo {author} {\bibfnamefont
  {R.}~\bibnamefont {Forza}}, \bibinfo {author} {\bibfnamefont
  {M.}~\bibnamefont {Manfrin}}, \ and\ \bibinfo {author} {\bibfnamefont
  {M.}~\bibnamefont {Onorato}},\ }\href@noop {} {\bibfield  {journal} {\bibinfo
   {journal} {Physical review letters}\ }\textbf {\bibinfo {volume} {118}},\
  \bibinfo {pages} {144503} (\bibinfo {year} {2017})}\BibitemShut {NoStop}%
\bibitem [{\citenamefont {Romero}\ \emph {et~al.}(2012)\citenamefont {Romero},
  \citenamefont {Melville},\ and\ \citenamefont {Kleiss}}]{romero2012jpo}%
  \BibitemOpen
  \bibfield  {author} {\bibinfo {author} {\bibfnamefont {L.}~\bibnamefont
  {Romero}}, \bibinfo {author} {\bibfnamefont {W.~K.}\ \bibnamefont
  {Melville}}, \ and\ \bibinfo {author} {\bibfnamefont {J.~M.}\ \bibnamefont
  {Kleiss}},\ }\href {\doibase 10.1175/JPO-D-11-072.1} {\bibfield  {journal}
  {\bibinfo  {journal} {Journal of Physical Oceanography}\ }\textbf {\bibinfo
  {volume} {42}},\ \bibinfo {pages} {1421} (\bibinfo {year}
  {2012})}\BibitemShut {NoStop}%
\bibitem [{\citenamefont {Lenain}\ and\ \citenamefont
  {Melville}(2017)}]{lenain2017jpo}%
  \BibitemOpen
  \bibfield  {author} {\bibinfo {author} {\bibfnamefont {L.}~\bibnamefont
  {Lenain}}\ and\ \bibinfo {author} {\bibfnamefont {W.~K.}\ \bibnamefont
  {Melville}},\ }\href {\doibase 10.1175/JPO-D-17-0017.1} {\bibfield  {journal}
  {\bibinfo  {journal} {Journal of Physical Oceanography}\ }\textbf {\bibinfo
  {volume} {47}},\ \bibinfo {pages} {2123} (\bibinfo {year}
  {2017})}\BibitemShut {NoStop}%
\bibitem [{\citenamefont {Deike}\ \emph {et~al.}(2014)\citenamefont {Deike},
  \citenamefont {Berhanu},\ and\ \citenamefont {Falcon}}]{deike2014pre}%
  \BibitemOpen
  \bibfield  {author} {\bibinfo {author} {\bibfnamefont {L.}~\bibnamefont
  {Deike}}, \bibinfo {author} {\bibfnamefont {M.}~\bibnamefont {Berhanu}}, \
  and\ \bibinfo {author} {\bibfnamefont {E.}~\bibnamefont {Falcon}},\ }\href
  {\doibase 10.1103/PhysRevE.89.023003} {\bibfield  {journal} {\bibinfo
  {journal} {Phys. Rev. E}\ }\textbf {\bibinfo {volume} {89}},\ \bibinfo
  {pages} {023003} (\bibinfo {year} {2014})}\BibitemShut {NoStop}%
\bibitem [{\citenamefont {Deike}\ \emph {et~al.}(2015)\citenamefont {Deike},
  \citenamefont {Miquel}, \citenamefont {Guti{\'e}rrez}, \citenamefont {Jamin},
  \citenamefont {Semin}, \citenamefont {Berhanu}, \citenamefont {Falcon},\ and\
  \citenamefont {Bonnefoy}}]{deike2015role}%
  \BibitemOpen
  \bibfield  {author} {\bibinfo {author} {\bibfnamefont {L.}~\bibnamefont
  {Deike}}, \bibinfo {author} {\bibfnamefont {B.}~\bibnamefont {Miquel}},
  \bibinfo {author} {\bibfnamefont {P.}~\bibnamefont {Guti{\'e}rrez}}, \bibinfo
  {author} {\bibfnamefont {T.}~\bibnamefont {Jamin}}, \bibinfo {author}
  {\bibfnamefont {B.}~\bibnamefont {Semin}}, \bibinfo {author} {\bibfnamefont
  {M.}~\bibnamefont {Berhanu}}, \bibinfo {author} {\bibfnamefont
  {E.}~\bibnamefont {Falcon}}, \ and\ \bibinfo {author} {\bibfnamefont
  {F.}~\bibnamefont {Bonnefoy}},\ }\href@noop {} {\bibfield  {journal}
  {\bibinfo  {journal} {Journal of Fluid Mechanics}\ }\textbf {\bibinfo
  {volume} {781}},\ \bibinfo {pages} {196} (\bibinfo {year}
  {2015})}\BibitemShut {NoStop}%
\bibitem [{\citenamefont {Waseda}\ \emph
  {et~al.}(2009{\natexlab{b}})\citenamefont {Waseda}, \citenamefont
  {Kinoshita},\ and\ \citenamefont {Tamura}}]{waseda2009interplay}%
  \BibitemOpen
  \bibfield  {author} {\bibinfo {author} {\bibfnamefont {T.}~\bibnamefont
  {Waseda}}, \bibinfo {author} {\bibfnamefont {T.}~\bibnamefont {Kinoshita}}, \
  and\ \bibinfo {author} {\bibfnamefont {H.}~\bibnamefont {Tamura}},\
  }\href@noop {} {\bibfield  {journal} {\bibinfo  {journal} {J. Phys.
  Oceanogr.}\ }\textbf {\bibinfo {volume} {39}},\ \bibinfo {pages} {2351}
  (\bibinfo {year} {2009}{\natexlab{b}})}\BibitemShut {NoStop}%
\bibitem [{\citenamefont {Fadaeiazar}\ \emph {et~al.}(2018)\citenamefont
  {Fadaeiazar}, \citenamefont {Alberello}, \citenamefont {Onorato},
  \citenamefont {Leontini}, \citenamefont {Frascoli}, \citenamefont {Waseda},\
  and\ \citenamefont {Toffoli}}]{FadaeiazarElmira2018Wtai}%
  \BibitemOpen
  \bibfield  {author} {\bibinfo {author} {\bibfnamefont {E.}~\bibnamefont
  {Fadaeiazar}}, \bibinfo {author} {\bibfnamefont {A.}~\bibnamefont
  {Alberello}}, \bibinfo {author} {\bibfnamefont {M.}~\bibnamefont {Onorato}},
  \bibinfo {author} {\bibfnamefont {J.}~\bibnamefont {Leontini}}, \bibinfo
  {author} {\bibfnamefont {F.}~\bibnamefont {Frascoli}}, \bibinfo {author}
  {\bibfnamefont {T.}~\bibnamefont {Waseda}}, \ and\ \bibinfo {author}
  {\bibfnamefont {A.}~\bibnamefont {Toffoli}},\ }\href@noop {} {\bibfield
  {journal} {\bibinfo  {journal} {Wave motion}\ }\textbf {\bibinfo {volume}
  {83}},\ \bibinfo {pages} {94} (\bibinfo {year} {2018})}\BibitemShut {NoStop}%
\bibitem [{\citenamefont {Dyachenko}\ and\ \citenamefont
  {Zakharov}(1994)}]{dyachenko1994free}%
  \BibitemOpen
  \bibfield  {author} {\bibinfo {author} {\bibfnamefont {A.~I.}\ \bibnamefont
  {Dyachenko}}\ and\ \bibinfo {author} {\bibfnamefont {V.~E.}\ \bibnamefont
  {Zakharov}},\ }\href@noop {} {\bibfield  {journal} {\bibinfo  {journal}
  {Physics Letters A}\ }\textbf {\bibinfo {volume} {190}},\ \bibinfo {pages}
  {144} (\bibinfo {year} {1994})}\BibitemShut {NoStop}%
\bibitem [{\citenamefont {Toffoli}\ \emph
  {et~al.}(2010{\natexlab{b}})\citenamefont {Toffoli}, \citenamefont {Babanin},
  \citenamefont {Onorato},\ and\ \citenamefont {Waseda}}]{toffoli10}%
  \BibitemOpen
  \bibfield  {author} {\bibinfo {author} {\bibfnamefont {A.}~\bibnamefont
  {Toffoli}}, \bibinfo {author} {\bibfnamefont {A.~V.}\ \bibnamefont
  {Babanin}}, \bibinfo {author} {\bibfnamefont {M.}~\bibnamefont {Onorato}}, \
  and\ \bibinfo {author} {\bibfnamefont {T.}~\bibnamefont {Waseda}},\ }\href
  {\doibase 10.1029/2009GL041771} {\bibfield  {journal} {\bibinfo  {journal}
  {Geophys. Res. Lett.}\ }\textbf {\bibinfo {volume} {37}},\ \bibinfo {pages}
  {L05603} (\bibinfo {year} {2010}{\natexlab{b}})}\BibitemShut {NoStop}%
\bibitem [{\citenamefont {Denissenko}\ \emph {et~al.}(2007)\citenamefont
  {Denissenko}, \citenamefont {Lukaschuk},\ and\ \citenamefont
  {Nazarenko}}]{denissenko2007prl}%
  \BibitemOpen
  \bibfield  {author} {\bibinfo {author} {\bibfnamefont {P.}~\bibnamefont
  {Denissenko}}, \bibinfo {author} {\bibfnamefont {S.}~\bibnamefont
  {Lukaschuk}}, \ and\ \bibinfo {author} {\bibfnamefont {S.}~\bibnamefont
  {Nazarenko}},\ }\href {\doibase 10.1103/PhysRevLett.99.014501} {\bibfield
  {journal} {\bibinfo  {journal} {Phys. Rev. Lett.}\ }\textbf {\bibinfo
  {volume} {99}},\ \bibinfo {pages} {014501} (\bibinfo {year}
  {2007})}\BibitemShut {NoStop}%
\bibitem [{\citenamefont {Falcon}\ \emph {et~al.}(2007)\citenamefont {Falcon},
  \citenamefont {Fauve},\ and\ \citenamefont
  {Laroche}}]{falcon2007observation}%
  \BibitemOpen
  \bibfield  {author} {\bibinfo {author} {\bibfnamefont {E.}~\bibnamefont
  {Falcon}}, \bibinfo {author} {\bibfnamefont {S.}~\bibnamefont {Fauve}}, \
  and\ \bibinfo {author} {\bibfnamefont {C.}~\bibnamefont {Laroche}},\
  }\href@noop {} {\bibfield  {journal} {\bibinfo  {journal} {Physical review
  letters}\ }\textbf {\bibinfo {volume} {98}},\ \bibinfo {pages} {154501}
  (\bibinfo {year} {2007})}\BibitemShut {NoStop}%
\bibitem [{\citenamefont {Falcon}\ \emph {et~al.}(2010)\citenamefont {Falcon},
  \citenamefont {Roux},\ and\ \citenamefont {Laroche}}]{falcon2010origin}%
  \BibitemOpen
  \bibfield  {author} {\bibinfo {author} {\bibfnamefont {E.}~\bibnamefont
  {Falcon}}, \bibinfo {author} {\bibfnamefont {S.~G.}\ \bibnamefont {Roux}}, \
  and\ \bibinfo {author} {\bibfnamefont {C.}~\bibnamefont {Laroche}},\
  }\href@noop {} {\bibfield  {journal} {\bibinfo  {journal} {EPL (Europhysics
  Letters)}\ }\textbf {\bibinfo {volume} {90}},\ \bibinfo {pages} {34005}
  (\bibinfo {year} {2010})}\BibitemShut {NoStop}%
\bibitem [{\citenamefont {Nazarenko}\ \emph {et~al.}(2010)\citenamefont
  {Nazarenko}, \citenamefont {Lukaschuk}, \citenamefont {McLelland},\ and\
  \citenamefont {Denissenko}}]{nazarenko2010statistics}%
  \BibitemOpen
  \bibfield  {author} {\bibinfo {author} {\bibfnamefont {S.}~\bibnamefont
  {Nazarenko}}, \bibinfo {author} {\bibfnamefont {S.}~\bibnamefont
  {Lukaschuk}}, \bibinfo {author} {\bibfnamefont {S.}~\bibnamefont
  {McLelland}}, \ and\ \bibinfo {author} {\bibfnamefont {P.}~\bibnamefont
  {Denissenko}},\ }\href@noop {} {\bibfield  {journal} {\bibinfo  {journal}
  {Journal of Fluid Mechanics}\ }\textbf {\bibinfo {volume} {642}},\ \bibinfo
  {pages} {395} (\bibinfo {year} {2010})}\BibitemShut {NoStop}%
\bibitem [{\citenamefont {Randoux}\ \emph {et~al.}(2014)\citenamefont
  {Randoux}, \citenamefont {Walczak}, \citenamefont {Onorato},\ and\
  \citenamefont {Suret}}]{randoux2014intermittency}%
  \BibitemOpen
  \bibfield  {author} {\bibinfo {author} {\bibfnamefont {S.}~\bibnamefont
  {Randoux}}, \bibinfo {author} {\bibfnamefont {P.}~\bibnamefont {Walczak}},
  \bibinfo {author} {\bibfnamefont {M.}~\bibnamefont {Onorato}}, \ and\
  \bibinfo {author} {\bibfnamefont {P.}~\bibnamefont {Suret}},\ }\href@noop {}
  {\bibfield  {journal} {\bibinfo  {journal} {Phys. Rev. Lett.}\ }\textbf
  {\bibinfo {volume} {113}},\ \bibinfo {pages} {113902} (\bibinfo {year}
  {2014})}\BibitemShut {NoStop}%
\bibitem [{\citenamefont {Connaughton}\ \emph {et~al.}(2003)\citenamefont
  {Connaughton}, \citenamefont {Nazarenko},\ and\ \citenamefont
  {Newell}}]{connaughton2003dimensional}%
  \BibitemOpen
  \bibfield  {author} {\bibinfo {author} {\bibfnamefont {C.}~\bibnamefont
  {Connaughton}}, \bibinfo {author} {\bibfnamefont {S.}~\bibnamefont
  {Nazarenko}}, \ and\ \bibinfo {author} {\bibfnamefont {A.~C.}\ \bibnamefont
  {Newell}},\ }\href@noop {} {\bibfield  {journal} {\bibinfo  {journal}
  {Physica D: Nonlinear Phenomena}\ }\textbf {\bibinfo {volume} {184}},\
  \bibinfo {pages} {86} (\bibinfo {year} {2003})}\BibitemShut {NoStop}%
\bibitem [{\citenamefont {Yokoyama}(2004)}]{yokoyama2004statistics}%
  \BibitemOpen
  \bibfield  {author} {\bibinfo {author} {\bibfnamefont {N.}~\bibnamefont
  {Yokoyama}},\ }\href@noop {} {\bibfield  {journal} {\bibinfo  {journal}
  {Journal of Fluid Mechanics}\ }\textbf {\bibinfo {volume} {501}},\ \bibinfo
  {pages} {169} (\bibinfo {year} {2004})}\BibitemShut {NoStop}%
\bibitem [{\citenamefont {Choi}\ \emph {et~al.}(2005)\citenamefont {Choi},
  \citenamefont {Lvov}, \citenamefont {Nazarenko},\ and\ \citenamefont
  {Pokorni}}]{choi2005anomalous}%
  \BibitemOpen
  \bibfield  {author} {\bibinfo {author} {\bibfnamefont {Y.}~\bibnamefont
  {Choi}}, \bibinfo {author} {\bibfnamefont {Y.~V.}\ \bibnamefont {Lvov}},
  \bibinfo {author} {\bibfnamefont {S.}~\bibnamefont {Nazarenko}}, \ and\
  \bibinfo {author} {\bibfnamefont {B.}~\bibnamefont {Pokorni}},\ }\href@noop
  {} {\bibfield  {journal} {\bibinfo  {journal} {Physics Letters A}\ }\textbf
  {\bibinfo {volume} {339}},\ \bibinfo {pages} {361} (\bibinfo {year}
  {2005})}\BibitemShut {NoStop}%
\bibitem [{\citenamefont {Hwang}\ \emph {et~al.}(2013)\citenamefont {Hwang},
  \citenamefont {Burrage}, \citenamefont {Wang},\ and\ \citenamefont
  {Wesson}}]{hwang2013ocean}%
  \BibitemOpen
  \bibfield  {author} {\bibinfo {author} {\bibfnamefont {P.~A.}\ \bibnamefont
  {Hwang}}, \bibinfo {author} {\bibfnamefont {D.~M.}\ \bibnamefont {Burrage}},
  \bibinfo {author} {\bibfnamefont {D.~W.}\ \bibnamefont {Wang}}, \ and\
  \bibinfo {author} {\bibfnamefont {J.~C.}\ \bibnamefont {Wesson}},\
  }\href@noop {} {\bibfield  {journal} {\bibinfo  {journal} {J. Atmos. Ocean
  Tech.}\ }\textbf {\bibinfo {volume} {30}},\ \bibinfo {pages} {2168} (\bibinfo
  {year} {2013})}\BibitemShut {NoStop}%
\bibitem [{\citenamefont {Waseda}\ \emph {et~al.}(2015)\citenamefont {Waseda},
  \citenamefont {Kinoshita}, \citenamefont {Cavaleri},\ and\ \citenamefont
  {Toffoli}}]{waseda2015third}%
  \BibitemOpen
  \bibfield  {author} {\bibinfo {author} {\bibfnamefont {T.}~\bibnamefont
  {Waseda}}, \bibinfo {author} {\bibfnamefont {T.}~\bibnamefont {Kinoshita}},
  \bibinfo {author} {\bibfnamefont {L.}~\bibnamefont {Cavaleri}}, \ and\
  \bibinfo {author} {\bibfnamefont {A.}~\bibnamefont {Toffoli}},\ }\href@noop
  {} {\bibfield  {journal} {\bibinfo  {journal} {Journal of Fluid Mechanics}\
  }\textbf {\bibinfo {volume} {784}},\ \bibinfo {pages} {51} (\bibinfo {year}
  {2015})}\BibitemShut {NoStop}%
\bibitem [{\citenamefont {Toffoli}\ \emph {et~al.}(2015)\citenamefont
  {Toffoli}, \citenamefont {Waseda}, \citenamefont {Houtani}, \citenamefont
  {Cavaleri}, \citenamefont {Greaves},\ and\ \citenamefont
  {Onorato}}]{toffoli2015rogue}%
  \BibitemOpen
  \bibfield  {author} {\bibinfo {author} {\bibfnamefont {A.}~\bibnamefont
  {Toffoli}}, \bibinfo {author} {\bibfnamefont {T.}~\bibnamefont {Waseda}},
  \bibinfo {author} {\bibfnamefont {H.}~\bibnamefont {Houtani}}, \bibinfo
  {author} {\bibfnamefont {L.}~\bibnamefont {Cavaleri}}, \bibinfo {author}
  {\bibfnamefont {D.}~\bibnamefont {Greaves}}, \ and\ \bibinfo {author}
  {\bibfnamefont {M.}~\bibnamefont {Onorato}},\ }\href@noop {} {\bibfield
  {journal} {\bibinfo  {journal} {Journal of Fluid Mechanics}\ }\textbf
  {\bibinfo {volume} {769}},\ \bibinfo {pages} {277} (\bibinfo {year}
  {2015})}\BibitemShut {NoStop}%
\bibitem [{\citenamefont {Komen}\ \emph {et~al.}(1994)\citenamefont {Komen},
  \citenamefont {Cavaleri}, \citenamefont {Donelan}, \citenamefont
  {Hasselmann}, \citenamefont {Hasselmann},\ and\ \citenamefont
  {Janssen}}]{komen94}%
  \BibitemOpen
  \bibfield  {author} {\bibinfo {author} {\bibfnamefont {G.}~\bibnamefont
  {Komen}}, \bibinfo {author} {\bibfnamefont {L.}~\bibnamefont {Cavaleri}},
  \bibinfo {author} {\bibfnamefont {M.}~\bibnamefont {Donelan}}, \bibinfo
  {author} {\bibfnamefont {K.}~\bibnamefont {Hasselmann}}, \bibinfo {author}
  {\bibfnamefont {H.}~\bibnamefont {Hasselmann}}, \ and\ \bibinfo {author}
  {\bibfnamefont {P.}~\bibnamefont {Janssen}},\ }\href@noop {} {\emph {\bibinfo
  {title} {Dynamics and modeling of ocean waves}}}\ (\bibinfo  {publisher}
  {Cambridge University Press},\ \bibinfo {address} {Cambridge},\ \bibinfo
  {year} {1994})\BibitemShut {NoStop}%
\bibitem [{\citenamefont {Toffoli}\ \emph
  {et~al.}(2010{\natexlab{c}})\citenamefont {Toffoli}, \citenamefont {Onorato},
  \citenamefont {Bitner-Gregersen},\ and\ \citenamefont
  {Monbaliu}}]{toffoli2010jgr}%
  \BibitemOpen
  \bibfield  {author} {\bibinfo {author} {\bibfnamefont {A.}~\bibnamefont
  {Toffoli}}, \bibinfo {author} {\bibfnamefont {M.}~\bibnamefont {Onorato}},
  \bibinfo {author} {\bibfnamefont {E.~M.}\ \bibnamefont {Bitner-Gregersen}}, \
  and\ \bibinfo {author} {\bibfnamefont {J.}~\bibnamefont {Monbaliu}},\ }\href
  {\doibase 10.1029/2009JC005495} {\bibfield  {journal} {\bibinfo  {journal}
  {Journal of Geophysical Research: Oceans}\ }\textbf {\bibinfo {volume} {115}}
  (\bibinfo {year} {2010}{\natexlab{c}}),\ 10.1029/2009JC005495}\BibitemShut
  {NoStop}%
\bibitem [{\citenamefont {Donelan}\ \emph {et~al.}(1996)\citenamefont
  {Donelan}, \citenamefont {Drennan},\ and\ \citenamefont
  {Magnusson}}]{donelan1996jpo}%
  \BibitemOpen
  \bibfield  {author} {\bibinfo {author} {\bibfnamefont {M.~A.}\ \bibnamefont
  {Donelan}}, \bibinfo {author} {\bibfnamefont {W.~M.}\ \bibnamefont
  {Drennan}}, \ and\ \bibinfo {author} {\bibfnamefont {A.~K.}\ \bibnamefont
  {Magnusson}},\ }\href {\doibase
  10.1175/1520-0485(1996)026<1901:NAOTDP>2.0.CO;2} {\bibfield  {journal}
  {\bibinfo  {journal} {Journal of Physical Oceanography}\ }\textbf {\bibinfo
  {volume} {26}},\ \bibinfo {pages} {1901} (\bibinfo {year}
  {1996})}\BibitemShut {NoStop}%
\bibitem [{\citenamefont {Holthuijsen}(2010)}]{holthuijsen2010waves}%
  \BibitemOpen
  \bibfield  {author} {\bibinfo {author} {\bibfnamefont {L.~H.}\ \bibnamefont
  {Holthuijsen}},\ }\href@noop {} {\emph {\bibinfo {title} {Waves in oceanic
  and coastal waters}}}\ (\bibinfo  {publisher} {Cambridge University Press},\
  \bibinfo {year} {2010})\BibitemShut {NoStop}%
\bibitem [{\citenamefont {Mitsuyasu}\ \emph {et~al.}(1975)\citenamefont
  {Mitsuyasu}, \citenamefont {Tasai}, \citenamefont {Suhara}, \citenamefont
  {Mizuno}, \citenamefont {Ohkusu}, \citenamefont {Honda},\ and\ \citenamefont
  {Rikiishi}}]{mitsuyasu1975jpo}%
  \BibitemOpen
  \bibfield  {author} {\bibinfo {author} {\bibfnamefont {H.}~\bibnamefont
  {Mitsuyasu}}, \bibinfo {author} {\bibfnamefont {F.}~\bibnamefont {Tasai}},
  \bibinfo {author} {\bibfnamefont {T.}~\bibnamefont {Suhara}}, \bibinfo
  {author} {\bibfnamefont {S.}~\bibnamefont {Mizuno}}, \bibinfo {author}
  {\bibfnamefont {M.}~\bibnamefont {Ohkusu}}, \bibinfo {author} {\bibfnamefont
  {T.}~\bibnamefont {Honda}}, \ and\ \bibinfo {author} {\bibfnamefont
  {K.}~\bibnamefont {Rikiishi}},\ }\href {\doibase
  10.1175/1520-0485(1975)005<0750:OOTDSO>2.0.CO;2} {\bibfield  {journal}
  {\bibinfo  {journal} {Journal of Physical Oceanography}\ }\textbf {\bibinfo
  {volume} {5}},\ \bibinfo {pages} {750} (\bibinfo {year} {1975})}\BibitemShut
  {NoStop}%
\bibitem [{\citenamefont {Goda}(2000)}]{goda2010}%
  \BibitemOpen
  \bibfield  {author} {\bibinfo {author} {\bibfnamefont {Y.}~\bibnamefont
  {Goda}},\ }\href {\doibase 10.1142/3587} {\emph {\bibinfo {title} {Random
  Seas and Design of Maritime Structures}}},\ Vol.~\bibinfo {volume} {33}\
  (\bibinfo {year} {2000})\BibitemShut {NoStop}%
\bibitem [{\citenamefont {Janssen}(2014)}]{janssen2014random}%
  \BibitemOpen
  \bibfield  {author} {\bibinfo {author} {\bibfnamefont {P.~A. E.~M.}\
  \bibnamefont {Janssen}},\ }\href@noop {} {\bibfield  {journal} {\bibinfo
  {journal} {J. Fluid Mech.}\ }\textbf {\bibinfo {volume} {759}},\ \bibinfo
  {pages} {236} (\bibinfo {year} {2014})}\BibitemShut {NoStop}%
\bibitem [{\citenamefont {El~Koussaifi}\ \emph {et~al.}(2018)\citenamefont
  {El~Koussaifi}, \citenamefont {Tikan}, \citenamefont {Toffoli}, \citenamefont
  {Randoux}, \citenamefont {Suret},\ and\ \citenamefont
  {Onorato}}]{el2018spontaneous}%
  \BibitemOpen
  \bibfield  {author} {\bibinfo {author} {\bibfnamefont {R.}~\bibnamefont
  {El~Koussaifi}}, \bibinfo {author} {\bibfnamefont {A.}~\bibnamefont {Tikan}},
  \bibinfo {author} {\bibfnamefont {A.}~\bibnamefont {Toffoli}}, \bibinfo
  {author} {\bibfnamefont {S.}~\bibnamefont {Randoux}}, \bibinfo {author}
  {\bibfnamefont {P.}~\bibnamefont {Suret}}, \ and\ \bibinfo {author}
  {\bibfnamefont {M.}~\bibnamefont {Onorato}},\ }\href@noop {} {\bibfield
  {journal} {\bibinfo  {journal} {Phys. Rev. E}\ }\textbf {\bibinfo {volume}
  {97}},\ \bibinfo {pages} {012208} (\bibinfo {year} {2018})}\BibitemShut
  {NoStop}%
\bibitem [{\citenamefont {Alberello}\ \emph {et~al.}(2019)\citenamefont
  {Alberello}, \citenamefont {Onorato}, \citenamefont {Frascoli},\ and\
  \citenamefont {Toffoli}}]{alberello2019observation}%
  \BibitemOpen
  \bibfield  {author} {\bibinfo {author} {\bibfnamefont {A.}~\bibnamefont
  {Alberello}}, \bibinfo {author} {\bibfnamefont {M.}~\bibnamefont {Onorato}},
  \bibinfo {author} {\bibfnamefont {F.}~\bibnamefont {Frascoli}}, \ and\
  \bibinfo {author} {\bibfnamefont {A.}~\bibnamefont {Toffoli}},\ }\href@noop
  {} {\bibfield  {journal} {\bibinfo  {journal} {Wave Motion}\ }\textbf
  {\bibinfo {volume} {84}},\ \bibinfo {pages} {81} (\bibinfo {year}
  {2019})}\BibitemShut {NoStop}%
\end{thebibliography}%
    \bibliographystyle {apsrev4-1}
        
\end{document}